\documentstyle[prd,aps,preprint,epsf]{revtex}
\def\li{\mbox{\rm Li$_2$}}
\def\al{\alpha}
\def\be{\beta}
\def\ga{\gamma}
\def\de{\delta}

\def\ep{\varepsilon}

\def\dot{\!\cdot\!}
\def\ds{\displaystyle}
\def\pr{\prime}
\def\epf#1#2#3#4{\varepsilon_{#1#2#3#4}}
\def\sla#1{#1\hspace*{-5pt}/}
\tighten
\begin{document}
\draft
\title{\hfill {\small DOE-ER-40757-118} \\
\hfill {\small UTEXAS-HEP-98-17} \\
\hfill  {\small MSUHEP-80930} \\
\hfill \\
Radiative corrections to the decays $\bbox{K^0_L\to e^+e^-}$ and
$\bbox{K^0_L\to\mu^+\mu^-}$}
\author{Duane A. Dicus} 
\address{Center for Particle Physics and Department of Physics\\ 
University of Texas, Austin, Texas 78712 }
\author{Wayne W. Repko} 
\address{Department of Physics and Astronomy, Michigan State University \\ 
East Lansing, Michigan 48824}
\date{\today}
\maketitle

\begin{abstract}
We calculate the rates and lepton ($\ell$) invariant mass distributions for 
decays of the form $0^{-+}\to \ell^+\ell^-\gamma$, which are important radiative
corrections to the purely leptonic decays $0^{-+}\to \ell^+\ell^-$. Our approach
uses the loop diagrams which arise by including the two photon intermediate
state and we retain the imaginary parts of the loops - a radiative extension of 
the `unitarity bound' for the process. These results are compared which those 
obtained using a model in which the meson couples directly to the leptons. 
\end{abstract}

\section{Introduction} \label{sec:1}
Recently, a few electron events in the channel $K^0_L\rightarrow e^+ e^-$,
where $K^0_L$ is the  long lived neutral $K$ meson, have been observed 
\cite{rit}.  The same experiment sees thousands of muon events, 
$K^0_L\to\mu^+ \mu^-$. One of the experimental acceptance conditions is that 
the invariant mass of the two leptons be within a few MeV of the $K$ mass.  
This means events are lost if a photon of sufficient energy is also emitted, 
{\em i.e.}\,$K^0_L\to\ell^+\ell^- \gamma$,\,with $\ell=e$ or $\mu$.  

Our purpose is to estimate the size of this radiative correction by calculating
\begin{equation}\label{intrate}
\frac{1}{\Gamma_0} \int^{(M-\Delta)^2}_{4m^2} ds \frac{d\Gamma}{ds}\,.
\end{equation}
Here $\Gamma$ is the rate for $K^0_L\to \ell^+\ell^- \gamma$, $\Gamma_0$ is the
rate for $K^0_L \to\ell^+\ell^-$ and $s$ in the square of the lepton
invariant mass, $s=(p+p')^2$, where $p$ and $p'$ are the momenta of the lepton 
and antilepton.  The $K$ mass is denoted by $M$, the lepton mass by $m$ and
$\Delta$ is an experimental parameter. In the next Section, we examine a tree
level model which treats the meson-lepton interaction as a point coupling. 
Section III contains results for a model in which the meson-lepton point 
interaction is replaced by a loop diagram with two photons in the intermediate 
state. We conclude with a comparison of the two approaches. Corrections to
$K^0_L\to\pi^+\pi^-\gamma$ are presented in an Appendix.

\section{Radiative corrections in the tree approximation}

The differential decay width $d\Gamma/ds$ has been calculated in a simple model
\cite{berg} where
the meson-lepton coupling is taken to be a pseudoscalar interaction with an 
effective coupling constant $g$. This leads to a transition amplitude $T$ of the
form
\begin{equation}\label{point}
T=ge\left(\bar u(p)\sla\ep\frac{(m + \sla{p} + \sla{k})}{2p\dot k}\gamma_5v(p') 
+ \bar{u}(p)\ga_5\frac{(m - \sla{p}' - \sla{k})}{2p'\dot k}\sla\ep v(p')\right)
\,,
\end{equation}
as illustrated in Fig.\,1. Here, $\ep$ is the photon polarization vector and $k$
is the photon momentum.
Unfortunately the result is given only in the limit $M\gg m$, a condition
that is clearly not satisfied for $K$ decay into muons. The extension to
include terms of all orders in $m^2/M^2$ is straightforward and we find that 
Eq.\,(\ref{point}) gives a differential decay width for $K_L^0\to
\ell^+\ell^-\gamma$ of the form
\begin{equation}\label{diffrate}
\frac{1}{\Gamma_0}\,\frac{d\Gamma}{ds}= \frac{\alpha}{\pi M^4}
\,\frac{1}{\sqrt{1 - 4m^2/M^2}}
\frac{1}{M^2-s}\;\left[(M^4+s^2-4M^2m^2)\ln\left(\frac{1+v}{1-v}\right) -
2M^2sv\right]
\end{equation}
where $v=\sqrt{1 - 4m^2/s}$. This is a slight extension of Bergstr\"om's 
expression\cite{berg} by the terms proportional to $m$ in the numerator and 
denominator.

In an attempt to estimate the model dependence of our corrections, we will
compare our results with ``model independent" corrections given by keeping
only the universal soft bremsstrahlung correction terms of the form
\begin{equation}
T\to ge\bar u(p)\gamma_5v(p')\left(\frac{p\dot\ep}{p\dot k} -
\frac{p^{\pr}\dot\ep}{p'\dot k}\right)\,.
\end{equation}
In this case the contents of
the square brackets in Eq.\,(\ref{diffrate}) are replaced by
\begin{equation}\label{polerate}
(2s^2 - 4sm^2)\ln\left(\frac{1+v}{1-v}\right) - 2vs^2.
\end{equation}

For the expressions (\ref{diffrate}) and (\ref{polerate}), the integral in 
Eq.\,(\ref{intrate}) can be evaluated analytically to
determine the fraction of lepton pairs missed. Using Eq.\,(\ref{diffrate}) we 
get
\begin{equation}\label{int}
\frac{1}{\Gamma_0}\int^{(M-\Delta)^2}_{4m^2} ds\,\frac{d\Gamma}{ds} =
\frac{\alpha}{\pi}\frac{1}{\sqrt{1-\ep^2}}\, F(\delta,\ep)
\end{equation}
where
$$ \delta = \frac{2\Delta}{M} -\frac{\Delta^2}{M^2}\, ,\;
\ep=\frac{2m}{M}$$
and
\begin{eqnarray}\label{F}
F(\de,\ep) & = & \left[-2(2 - \ep^2)\left(\ln\de + 2\ln(1 + \be_-^2)\right) -
2\left(1 - \de + \case{1}{2}(1 - \de)^2\right) + 2(1 - \sqrt{1 - \ep^2})^2
\right. \nonumber \\
&   &\left.+ \ep^2(1 + \case{3}{8}\ep^2)\right]\ln\al_+
+ (2 - \ep^2)\left[-\li\left(\frac{\al_+^2}{\be_+^2}\right) + 
\li\left(\frac{\be_-^2}{\al_+^2}\right)\right] \nonumber \\
&   & -2\sqrt{1 - \ep^2}\ln\left(\frac{1 - \be_-^2/\al_+^2}{1 -
\al_+^2/\be_+^2}\right) + (\case{13}{4} - \case{1}{4}\de + \case{3}{8}\ep^2)
\sqrt{1 - \de}\sqrt{1 - \de - \ep^2}\,,
\end{eqnarray}
with
\begin{eqnarray}
\al_+ & = & \frac{\ds\sqrt{1 - \de}}{\ep} + \frac{\ds\sqrt{1 - \de -
\ep^2}}{\ep}\,, \\
\be_{\pm} & = & \frac{1}{\ep} \pm \frac{\ds\sqrt{1 - \ep^2}}{\ep}\,.
\end{eqnarray}
$\li(x)$ is a Spence function or dilogarithm defined as
$$
\li(x)=-\int_0^x \,\frac{dt}{t}\ln(1-t) \,.
$$
Numerical values of $\li(x)$ can easily be obtained using Maple or Mathematica.

For expression (\ref{polerate}) we again have Eq.\,(\ref{int}) where now
\begin{eqnarray}\label{Fpole}
F_{\rm pole}(\de,\ep) & = &\left[-2(2 - \ep^2)\left(\ln\de + 2\ln(1 + \be_-^2)
\right) - 2\left((2 - \ep^2)(1 - \de) + (1 - \de)^2\right)\right.
\nonumber \\
&   &\left.+ 2(1 - \sqrt{1 - \ep^2})^2 + \ep^2(2 - \case{3}{4}\ep^2)\right]
\ln{\al_+} +  (2 - \ep^2)\left[-\li\left(\frac{\al_+^2}{\be_+^2}\right) +
\li\left(\frac{\be_-^2}{\al_+^2}\right)\right] \nonumber \\ 
&   &-2\sqrt{1 - \ep^2}\ln\left(\frac{1 - \be_-^2/\al_+^2}{1 -
\al_+^2/\be_+^2}\right) + (\case{11}{2} - \case{3}{2}\de - \case{3}{4}\ep^2)
\sqrt{1 - \de}\sqrt{1 - \de - \ep^2}\,.
\end{eqnarray}

\section{Radiative corrections using a one-loop model}

All the experimental results announced so far \cite{mumu} find the rate for 
$K^0_L\to\mu^+\mu^-$ to be near the theoretical lower limit given by 
multiplying the rate for $K^0_L\to \gamma\gamma$ by the rate for 
$\gamma\gamma\to \mu^+\mu^-$
\begin{equation}\label{kggmumu}
\Gamma=\Gamma (K^0_L\to \gamma\gamma)\frac{\alpha^2}{2\beta}\left[\frac{m}{M}\,
\ln\left(\frac{1+\beta}{1-\beta}\right)\right]^2\,,
\end{equation}
where
$$
\beta=\sqrt{1 - 4m^2/M^2}.
$$
The rate for $K_L^0\to e^+e^-$ as given in \cite{rit} is larger than the 
unitarity bound given by Eq.\,(\ref{kggmumu}) and consistent with predictions 
from chiral perturbation theory \cite{chlag}.  Nevertheless, it seems 
reasonable, in attempting to extend the calculation of 
Eq.\,(\ref{intrate}) beyond the result obtained using Eq.\,(\ref{point}), to
calculate the absorptive part of  $K^0_L\rightarrow\ell^+\ell^-\gamma$ 
diagrams shown in Fig.\,2. In particular the box diagram takes us beyond simple
bremsstrahlung off external legs. Like Eq.\,(\ref{kggmumu}), the bremsstrahlung
pole terms of Fig.\,(2), which vary as $1/\omega$, where $\omega$ is the photon 
energy, contain a factor of the lepton mass. It has been suggested that the 
terms which vary as $\omega$ to a positive power might not include a lepton 
mass factor and could thus be anomalously large \cite{bed}.

To evaluate the diagrams of Fig.\,2, we take the $K$-photon-photon effective
Lagrangian to be 
\begin{equation}\label{kgg}
A_{\gamma\gamma}\phi\, F^{\mu\nu}\tilde F_{\mu\nu}\,,
\end{equation}
where $F^{\mu\nu}$ is the photon field tensor, $\tilde F_{\mu\nu}$ is its dual,
$\phi$ is the $K_L^0$ field and $A_{\gamma\gamma}$ is a constant. This leads to
the vertex function 
\begin{equation}
\Gamma_{\mu\nu}(k,k') = 2A_{\ga\ga}\,\epf\mu\nu\al\be\, k^{\al}k'^{\be}\,,
\end{equation}
where $k$ and $k'$ are the photon momenta. In general, this expression can also 
include a form factor which depends on $k^2$ and $k'^2$. The implications 
of including this additional factor are discussed below.

In this case, the expression for $d\Gamma/ds$ is very
complicated and we will not attempt to write it out. The box graph involves 
integrals of the form
\begin{equation}\label{boxint}
\int d^4q\;\frac{q^\mu q^\nu q^\alpha\,,\, q^\mu q^\nu\,,\,
q^\mu}{q^2\left[(q+p)^2-m^2\right]\,\left[(q+p+k)^2-m^2\right]\, (q+ P)^2}.
\end{equation}
The triangle graphs involve similar integrals with one or two factors of $q$ in 
the numerator and either the second or third factor in the denominator of 
Eq.\,(\ref{boxint}) omitted.  These integrals can be expanded in terms of the 
external 
momenta as outlined in the appendix of Passarino and Veltman \cite{pv}. The 
momentum expansion and its scalar coefficients are given by a
computer code called LOOP \cite{'thv,dk} which is a slight modification of a 
code written by Veltman called FORM Factor.  Within this code the integrals 
are evaluated in terms of Spence functions as defined above and these 
functions are then evaluated numerically.

Once the amplitudes are determined they are squared and summed over spin,
including the photon polarization, in the usual way. To check for errors we 
replace the photon polarization by its momentum and look for gauge invariance. 
This is our only real check but it is a very powerful one since gauge 
invariance requires a delicate cancellation among the three diagrams
\cite{brute}.

The real part of the amplitudes diverges because the effective coupling
Eq.\,(\ref{kgg}) has dimension 5.  The absorptive part has several 
contributions: $K\to\gamma\gamma$ followed by $\gamma\gamma\rightarrow\ell^+
\ell^-\gamma$ 
as well as $K\to\gamma\ell^+\ell^-$ followed by $\gamma\ell\to\gamma\ell$.
This is illustrated by the cut diagrams of Fig.\,3. In the first of these
diagrams, the intermediate photons are on-shell ($k^2 = k'^2 = 0$), which is
equivalent to our assumption that $A_{\ga\ga}$ is constant. In the second, one
of the photons is virtual, and the effective coupling has the general form
$A_{\ga\ga^*} = A_{\gamma\gamma}f(k^2)$, where $f(k^2)$ is a form factor
normalized to $f(0) = 1$ \cite{dalitz}. Our numerical calculation of the 
complete absorptive part cannot separate these contributions, so we have 
effectively assumed $A_{\gamma\gamma^*} = A_{\gamma\gamma}$ throughout. This
assumption is justified in the case of electrons, since the form factor
correction to the width of the Dalitz decay $K^0_L\to e^+e^-\ga$ is only a few 
percent due to the preference for low $e^+e^-$ invariant mass. For the muon 
case, the form factor correction to the Dalitz decay width is 20-25\% 
\cite{dalitz}, and there could be a discernable effect in the one loop 
contributions to $d\Gamma/ds$. 

\section{Results and conclusions}

In Figs.\,4, 5 and 6, we show the differential width obtained from the 
absorptive
part of the one-loop calculation as a function of the lepton invariant mass, 
$s$, for $K^0_L\to e^+e^-\gamma$, $K_L^0\to \mu^+\mu^-\gamma$, and
$\pi^0\to e^+e^-\gamma$.  For comparison we also plot Eqs.\,(\ref{diffrate}) 
and (\ref{polerate}). The result of the loop calculation is almost the same as 
Bergstr\"om's differential width, as modified by us to include the lepton mass, 
and both differ substantially from the ``model independent" width where only 
the $1/\omega $ terms are kept.

In Tables I, II, and III we give the integrated width, Eq.\,(1), for several 
values 
of $\Delta$. Again the result from the loop calculation is very similar to
that given by Eq.\,(\ref{F}) and quite different from that given by 
Eq.\,(\ref{Fpole}). For electrons the correction is large, but not anomalously 
so, and there is no indication that the diagrams of Fig.\,2 are not 
proportional to the lepton mass.

The moral would seem to be that, except for very small invariant masses,
the more complicated calculation of the loop model of Sec.\,III is
unnecessary and Eqs.\,(\ref{diffrate}) and (\ref{int}) are sufficient.
We have made no attempt to calculate the radiative corrections
within the acceptance bin, $(M-\Delta)^2 < s < M^2$.   For the model of
Sec.\,II Bergstr\"om \cite{berg} has given a complete expression for the 
correction from virtual photons.  This, together with Eq.\,(\ref{diffrate}), 
is all that is needed. To calculate the virtual corrections for the model of 
Sec.\,III is beyond the scope of this work.

\acknowledgements

  It is a pleasure to thank Jack Ritchie for drawing our attention
to this problem, for many discussions, and for preliminary versions
of Tables I and II and Figs.\,4 and 5.   We also thank Jack Smith
for providing us with many references, in particular Ref.\,\cite{chlag}. This 
work was supported in part by the U.S. Department of Energy under
Contract No. DE-FG013-93ER40757 and in part by the National Science Foundation
under Grant No. PHY-98-02439.

\appendix
\section*{}

In a model for the $K^0\to\pi^+\pi^-$ vertex similar to Eq.\,(\ref{point}), 
where the $K\pi\pi $ vertex is taken as a constant, the differential rate for
$K^0\rightarrow\pi^+\pi^-\gamma$ has only the $1/\omega$ terms and is therefore 
given by Eq.\,(\ref{polerate}) multiplied by $M^2/s$ to remove the spinor 
factor \cite{tau}. The integrated rate is given by Eq.\,(\ref{int}) with
\begin{eqnarray}\label{pipole}
F_{\rm pole}(\de,\ep) & = &\left[-2(2 - \ep^2)\left(\ln\de + 2\ln(1 + \be_-^2)
\right) - 4(1 - \de) + 2(1 - \sqrt{1 - \ep^2})^2\right.
\nonumber \\
&   &\left. + 2\ep^2\right]\ln{\al_+}
+  (2 - \ep^2)\left[-\li\left(\frac{\al_+^2}{\be_+^2}\right) +
\li\left(\frac{\be_-^2}{\al_+^2}\right)\right] \nonumber \\
&   &-2\sqrt{1 - \ep^2}\ln\left(\frac{1 - \be_-^2/\al_+^2}{1 -
\al_+^2/\be_+^2}\right) + 4\sqrt{1 - \de}\sqrt{1 - \de - \ep^2}\,.
\end{eqnarray}

For the four values of $\Delta$ used in $K^0_L\to\ell^+\ell^-\gamma$,\, 
$\Delta$ = 7.67, 5.67, 3.67, and 1.67 MeV the fractional radiative corrections 
are 0.0127, 0.0145, 0.0172 and 0.0223.

\begin{figure}[h]
\centerline{ \epsfbox{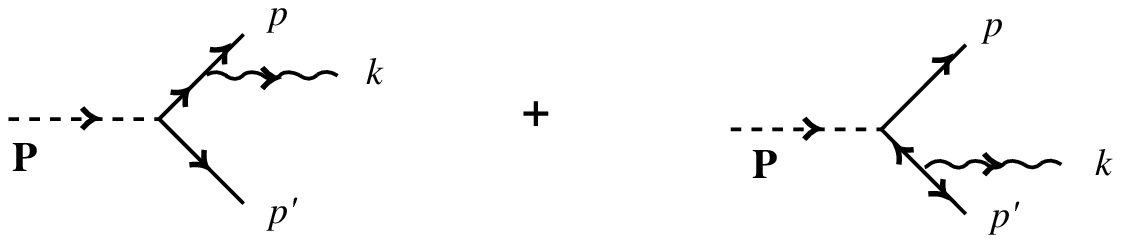}}
\vspace{20pt}
\caption{Diagrams for the radiative corrections to the tree model are shown.
A dashed line denotes a meson, a wavey line denotes a photon and a solid line
denotes a lepton.}
\end{figure}


\begin{figure}[h]
\centerline{ \epsfbox{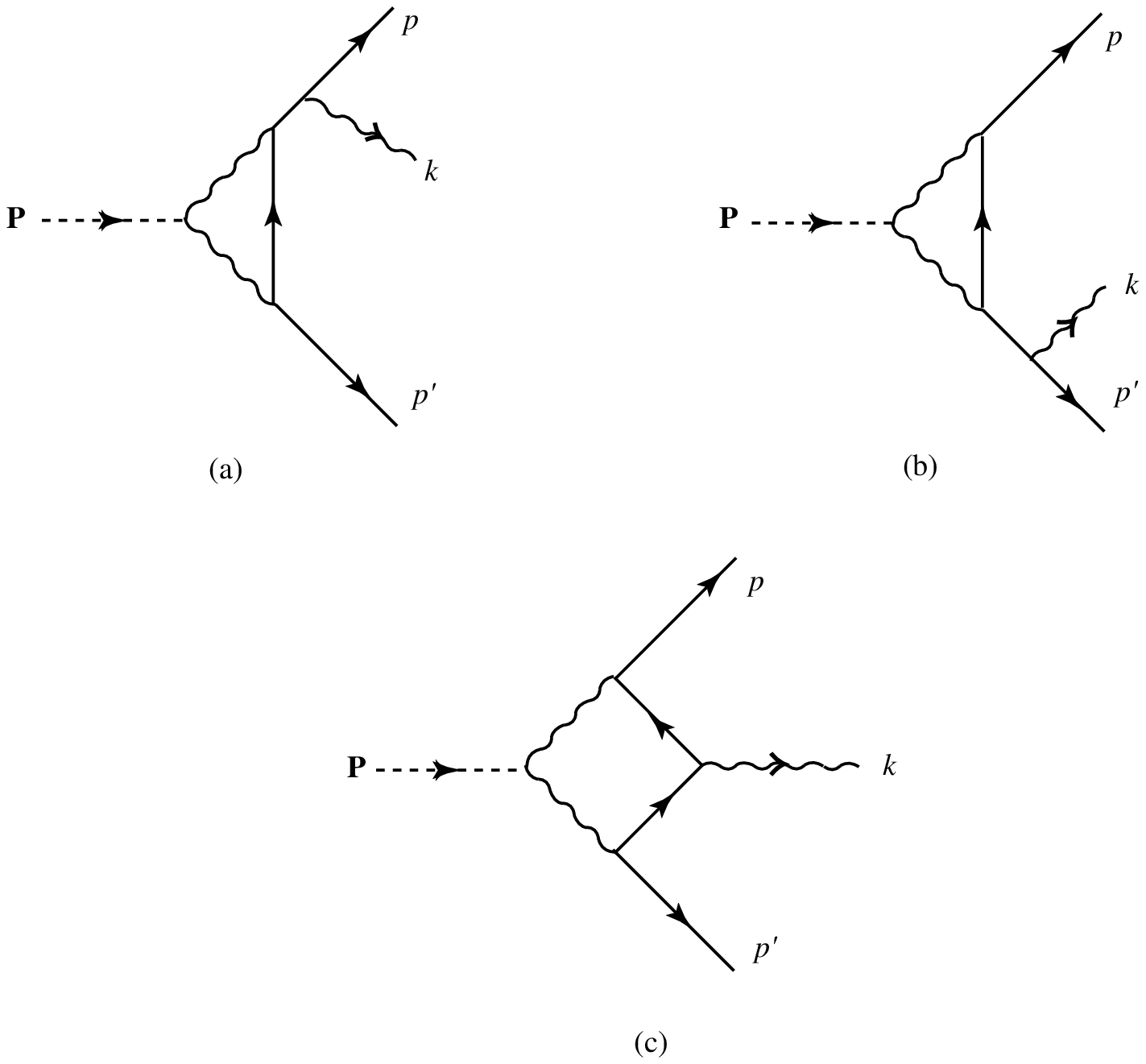}}
\vspace{20pt}
\caption{Diagrams for the radiative corrections to the one-loop model are shown.
A dashed line denotes a meson, a wavey line denotes a photon and a solid line
denotes a lepton.}
\end{figure}

\newpage

\begin{figure}[h]
\hspace{.2in}
\epsfysize=2.0in \epsfbox{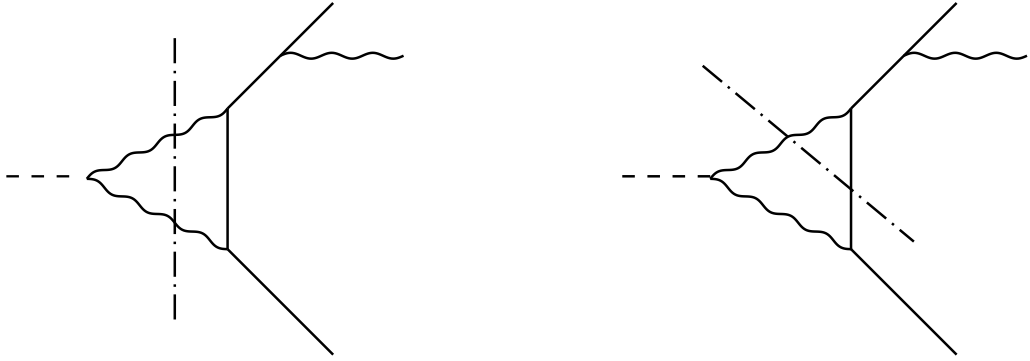}
\vspace{20pt}
\caption{Typical cut diagrams for the contributions to the absorptive part of 
the one-loop model are shown. The dot-dashed lines indicate the
propagators which are put on mass shell. The sum of these two diagrams
determines the imaginary part of the diagram in Fig.\,2\,(a). Figs.\,2\,(b)
and 2\,(c) have similar cuts.}
\end{figure}

\newpage

\begin{figure}[h]
\epsfysize=5.5in \epsffile{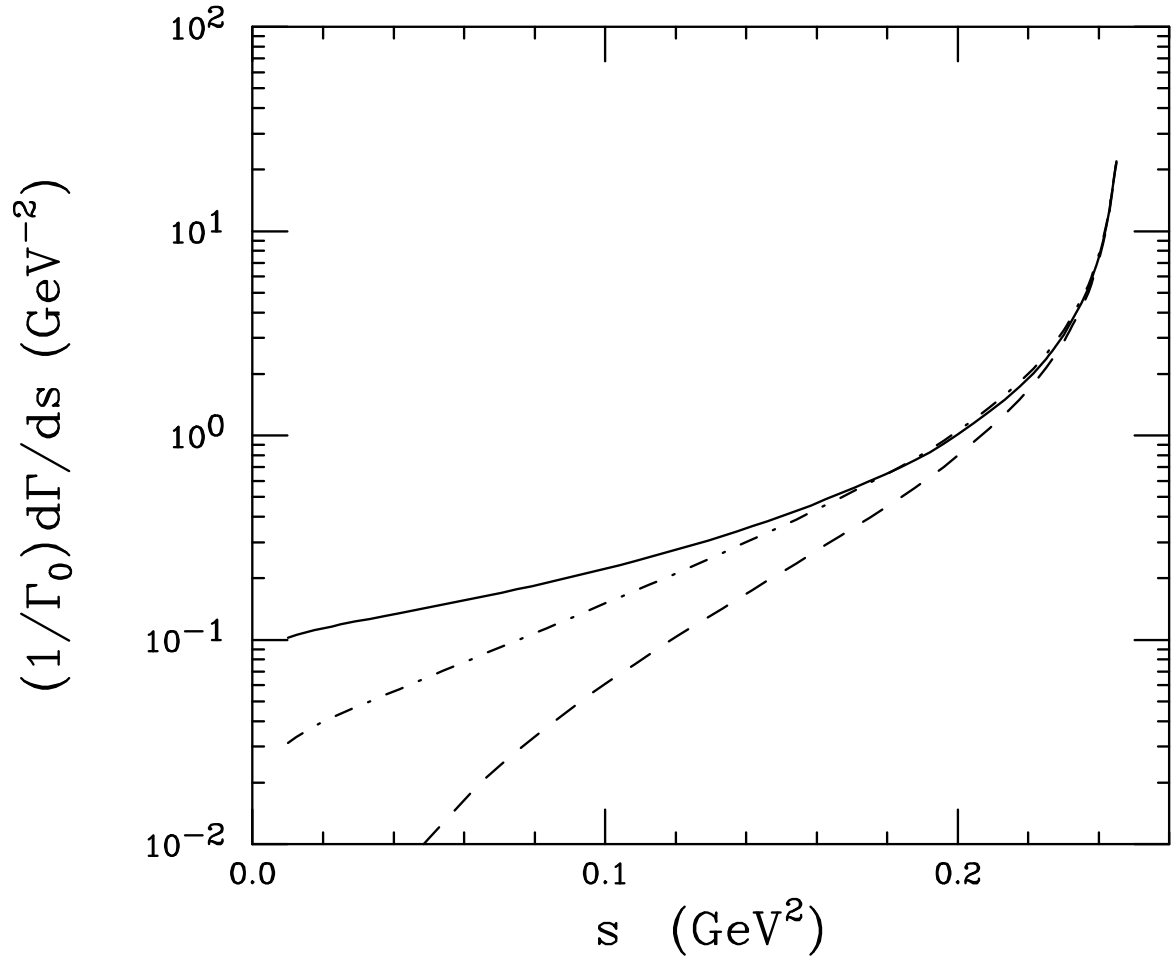}
\vspace{10pt}
\caption{The $e^+e^-$ invariant mass ($s$) distribution for the decay 
$K^0_L\protect\to e^+e^-\gamma$ is shown normalized to the decay width 
$\Gamma_0(K^0_L\protect\to e^+e^-)$. The solid line is the result of 
Ref.\protect\cite{berg}, the dashed line is the contribution of the 
$1/\omega$ poles, Eq.\,(\protect\ref{polerate}), and the dot-dash line is the 
result for the loop model of Sec.\,III.}
\end{figure}

\newpage

\begin{figure}[h]
\epsfysize=5.5in \epsffile{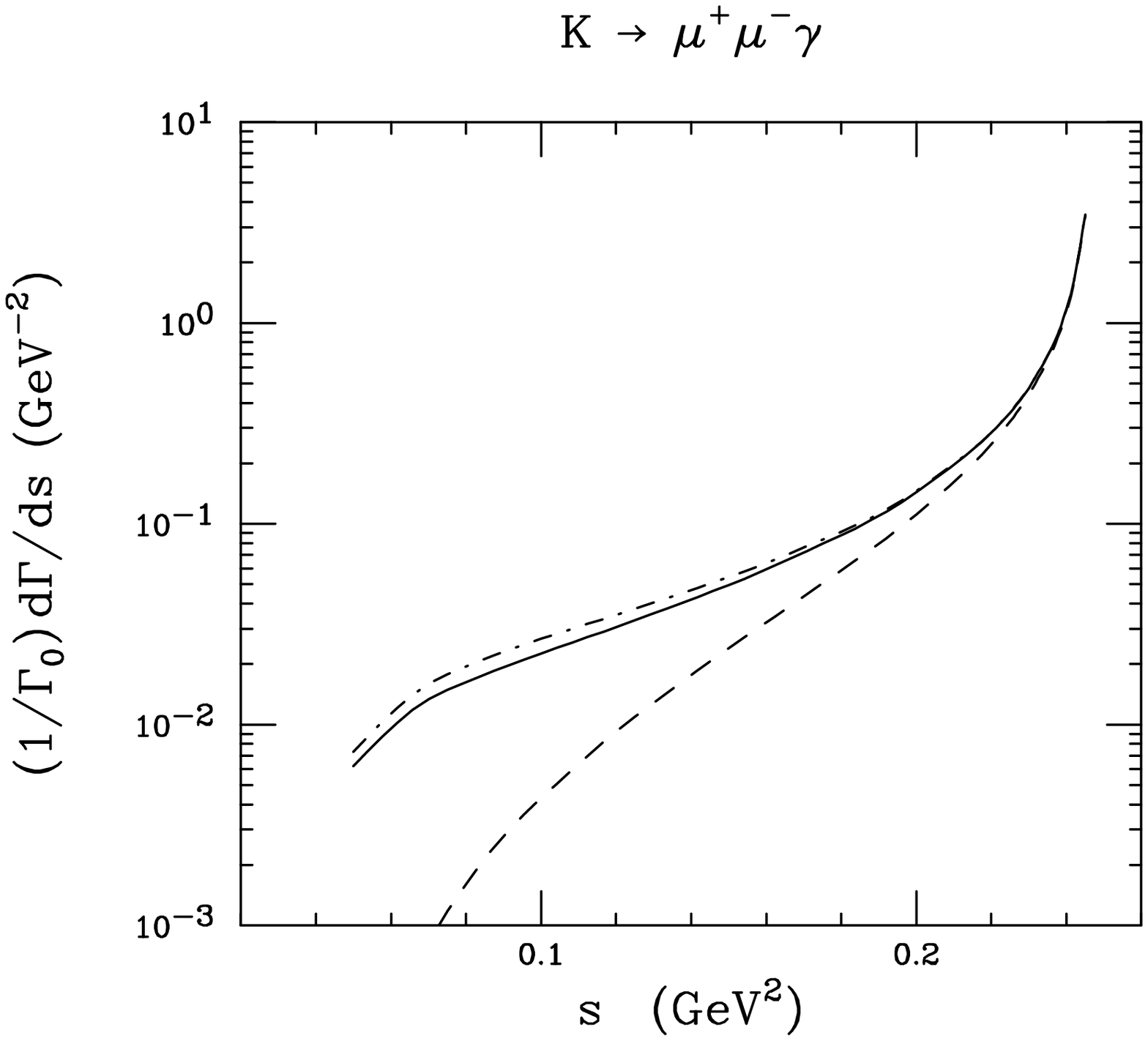}
\vspace{10pt}
\caption{Same as Fig.3 for $K\protect\to \mu^+\mu^-\gamma$.}
\end{figure}

\newpage

\begin{figure}[h]
\epsfysize=5.5in \epsffile{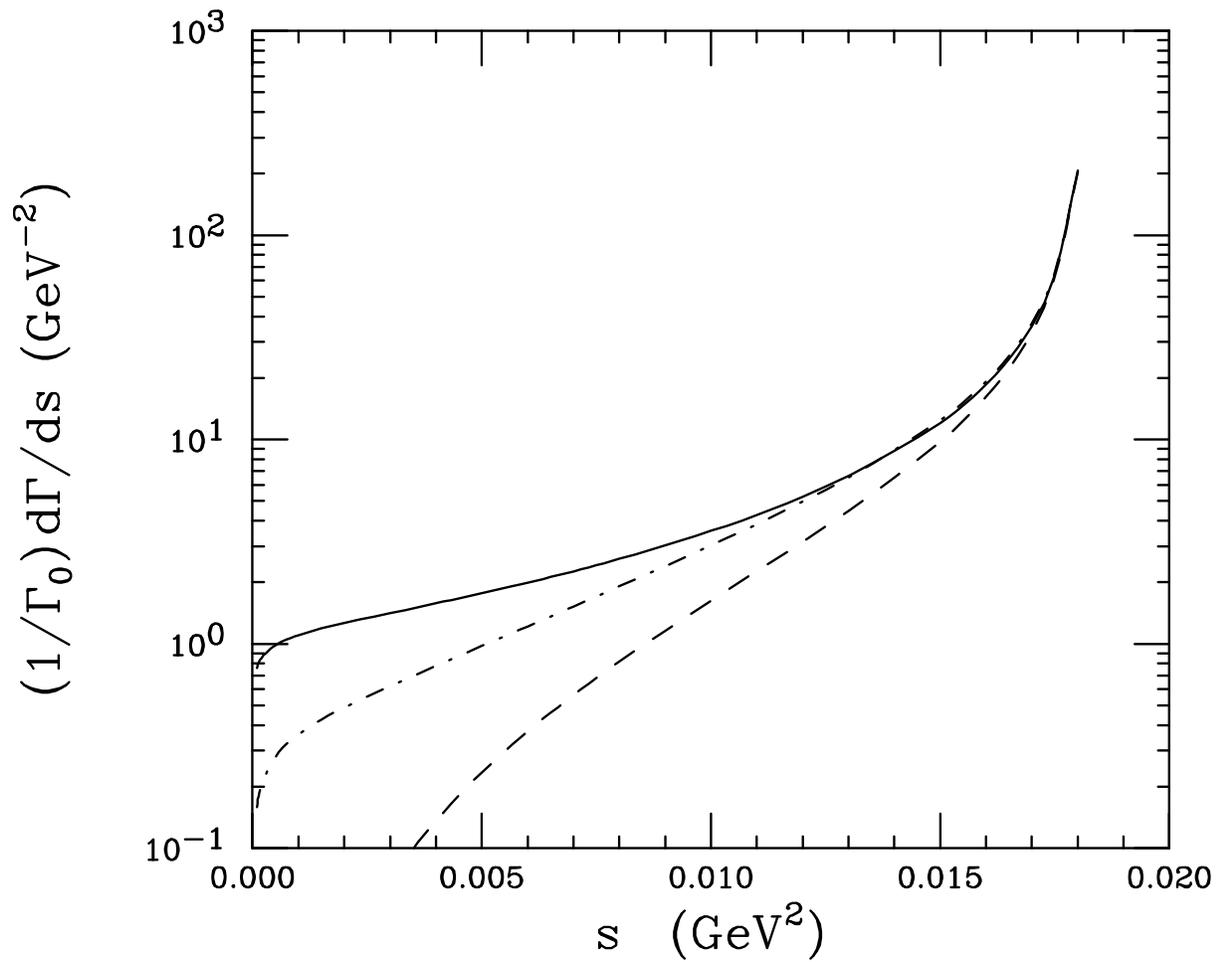}
\vspace{10pt}
\caption{Same as Fig.1 for $\pi\protect\to e^+e^-\gamma$.}
\end{figure}

\begin{table}
\begin{tabular}{cccc}
$\Delta$ (MeV) & Modified Ref.(2) & Loop & Model Indep. \\ \hline
7.64 & 0.161 & 0.154 & 0.120 \\
5.67 & 0.178 & 0.172 & 0.137 \\
3.67 & 0.203 & 0.197 & 0.161 \\
1.67 & 0.249 & 0.244 & 0.207         
\end{tabular}
\caption{The fractional radiative correction for $K^0_L\protect\to e^+e^-$,
as defined  by Eq.\,(\protect\ref{intrate}), for several values of the cutoff 
$\Delta$ is shown. The $K$ mass is taken to be 497.67 MeV.  The second 
column is given by Eq.\,(\protect\ref{F}), the third column is given by the 
model of Sec.\,III and the fourth column by Eq.\,(\protect\ref{Fpole}).}
\end{table}

\vspace{20pt}

\begin{table}
\begin{tabular}{cccc}
$\Delta$ (MeV) & Modified Ref.(2) & Loop & Model Indep.\\ 
\hline
7.67  & 0.0217  & 0.0224  & 0.0171   \\
5.67  & 0.0244  & 0.0251  & 0.0198   \\
3.67  & 0.0284  & 0.0290  & 0.0236   \\
1.67  & 0.0357  & 0.0363  & 0.0309           
\end{tabular}
\caption{Same as Table I for $K^0_L\protect\to \mu^+\mu^-$}
\end{table}

\vspace{20pt}

\begin{table}
\begin{tabular}{cccc}
$\Delta$ (MeV) & Modified Ref.(2) & Loop & Model Indep.  \\
\hline
4 &0.098 & 0.098  & $6.71\times 10^{-2}$           \\
3 &0.111 & 0.111  & $7.91\times 10^{-2}$          \\
2 &0.129 & 0.129  & $9.67\times 10^{-2}$          \\
1 &0.161 & 0.161  & 0.128                         
\end{tabular}
\caption{Same as Table I for $\pi^0\protect\to e^+e^-$. We use $m_{\pi} = 135$\,
MeV.}
\end{table}


\begin{thebibliography}{99}
\bibitem{brute} This problem is simple enough to do all this by brute force 
using Vermaseren's code FORM or Veltman's code SCHOONSHIP.
\end{thebibliography}
\end{document}